\documentclass{elsart}
\usepackage{natbib}
\usepackage{epsfig}
\begin{document}
\runauthor{Cicero, Caesar and Vergil}
\begin{frontmatter}
\title{
Distortions of Experimental
Muon Arrival Time Distributions of Extensive
Air Showers by the Observation Conditions
}
\author[Fzka]{R. Haeusler}
\author[Unika,Buch]{A.F. Badea}
\author[Fzka,Unihd]{H. Rebel\thanksref{Mailaddr}}
\author[Buch]{I.M. Brancus}
\author[Fzka]{J. Oehlschl\"ager}

\address[Fzka]{Forschungszentrum Karlsruhe, Institut f\"ur Kernphysik, 
76021 Karlsruhe, Germany.}
\address[Unika]{University of Karlsruhe, Institut f\"ur 
Experimentelle Kernphysik, 76021 Karlsruhe, Germany.}
\address[Buch]{National Institute of Physics and Nuclear Engineering, 
76900 Bucharest, Romania.}
\address[Unihd]{University of Heidelberg, Faculty of Physics and 
Astronomy, 69120 Heidelberg, Germany.}
\thanks[Mailaddr]{Corresponding author: rebel@ik3.fzk.de}
 
\begin{abstract}
Event-by-event measured arrival time distributions of
Extensive Air Shower (EAS) muons are affected and
distorted by various interrelated effects which originate
from the time resolution of the timing
detectors, from fluctuations of the reference time and the
number (multiplicity) of detected
muons spanning the arrival time distribution of the
individual EAS events. The origin of these
effects is discussed, and different correction procedures,
which involve detailed simulations,
are proposed and illustrated.
The discussed distortions are relevant for relatively small observation 
distances ($R_\mu$ $<$ 200 m) from the EAS core. Their significance   
decreases with increasing observation distance and increasing primary 
energies. Local arrival time distributions which refer to the observed arrival 
time of the first local muon prove to be less sensitive to the mass of the 
primary. This feature points to the necessity of arrival time measurements with 
additional information on the curvature of the EAS disk.
\end{abstract}
 
\end{frontmatter}


\section{Introduction}

The pioneering studies of the EAS disk structure in the early 1950`s by Bassi, Clark and Rossi 
[1] have been followed by increasing efforts (e.g. [2-8]) to detail the information,  up to  
recent measurements with modern  detector arrays [9-11] like KASCADE [12]. Especially for 
the EAS muon component [12-14] the basic interest arises from the question, how 
well does the temporal EAS structure reflects the longitudinal profile and how well do muon arrival time 
distributions map the muon production heights, in particular when observed at sufficiently 
large distances from the shower axis, where the path-length  effects dominate [16]. Via these 
features muon arrival time distributions should carry some information on the mass of the 
EAS primaries, unless the intrinsic fluctuations of the muon generation processes and 
limitations of the detector response do obscure the discrimination. Experimental studies with 
these aspects are a current subject of the KASCADE experiment [17-20], taking 
use of the
\begin{figure}[b]
\begin{center}
\epsfig{file=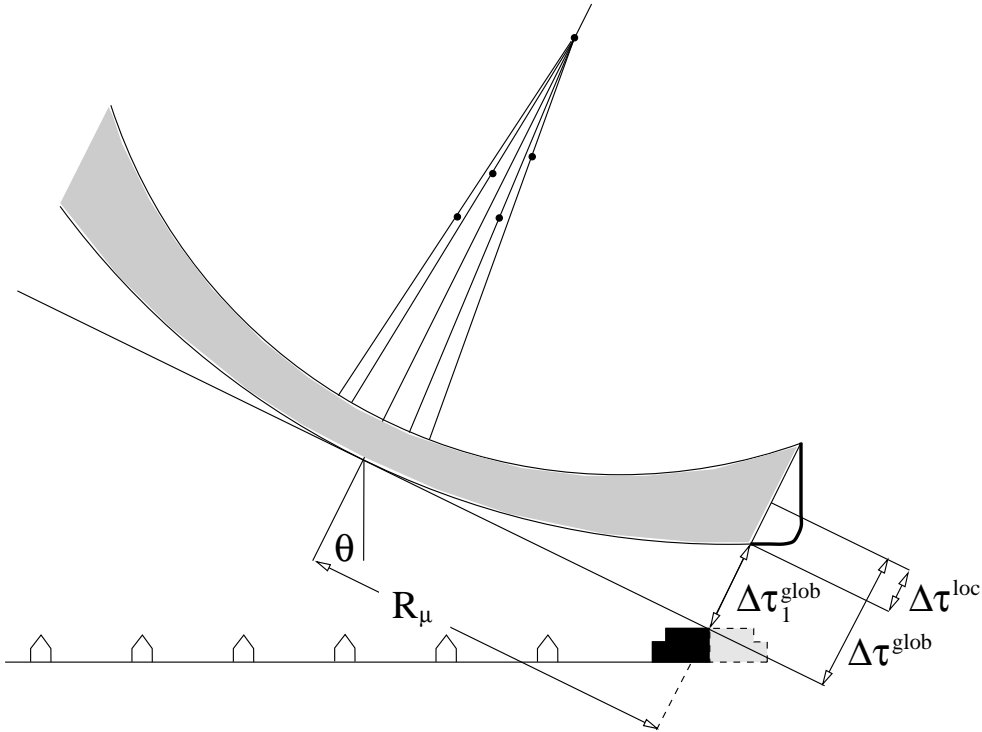, width=13.0cm}
\caption{
Characterisation of the EAS temporal structure by global and local arrival time 
parameters.
}
\end{center}
\end{figure} 
timing and muon detection facilities of the KASCADE Central Detector [21,22].
Experimental muon arrival time distributions have experienced various intriguing 
distortions which depend not only on the time resolution of the timing detectors, but also on 
the multiplicity, i.e. the number {\it n} of registered muons spanning the arrival time distribution of 
the single EAS event. Thus the shapes of the muon arrival time distribution are affected by the 
efficiency of the detection system registering the fluctuating number of EAS muons.
In this note we briefly discuss these multiplicity effects entangled with the response and the 
time resolution of the apparatus, and we suggest some procedures to 
reveal the basic time structure of the observed EAS.
Fig.~1 indicates the used terminology (see also [17]). Arrival times of muons, 
registered by the timing detectors at a certain distance $R_{\mu}$ from the 
shower axis have to refer to a defined zero time. For the reference zero the 
arrival time $\tau_c$ of the shower core in the detection plane could be used 
(global arrival times). The correction $\pm R_\mu tan\theta /c $ ($c$ - speed of
light) is applied to muon arrival times in order to eliminate the distorsions
due to the shower inclination.
However, there are often experimental difficulties to 
reconstruct this time-zero with sufficient accuracy.
Hence alternatively local arrival times are considered 
which refer to the arrival time of the 
first muon, registered locally. In event-by-event 
observations the individual EAS relative 
arrival time distributions (global and local) can be characterised by 
mean values $\Delta\tau_{mean}$ and by 
various quantiles like the median $\Delta\tau_{0.5}$, 
the first and the third quartiles 
$\Delta\tau_{0.25}$ and $\Delta\tau_{0.75}$, 
which describe different features of the
single distributions. The variation of the distributions 
of these quantiles, in particular of their
mean values and the variances, with the distance $R_\mu$ from 
the shower centre, we call 
the EAS time profile. In the case of global time parameters 
these describe the curvature of the 
shower disk and the shower thickness, while local quantities 
characterise only the structure of 
the shower disk.
The following discussion is mainly focused on the implications 
of observations of local time 
quantities and is based on Monte Carlo simulations of 
the EAS development, using the code 
CORSIKA [23] with the hadronic interaction model QGSJET [24] as generator.



\section{Fluctuations of the arrival time of the first muon}

The question to which extent the first detected particle represents 
the arrival time of the EAS front has been discussed by Villiers et.
al. [25].\\
Following these considerations it can be argued that the arrival time 
$\Delta\tau_{1}$ of the first muon, relative to a fictitious zero time, 
representing the muon front, approximated by
a sample with a large number $n$ of muons, its expectation 
value and fluctuations depend on
the particular value of the multiplicity $n$. Fig.~2 shows 
muon arrival time distributions
accumulated from many showers with the expectation values $\langle 
\Delta\tau_{1}(n)\rangle$ of the arrival time of
the first muon for subsamples of different multiplicities.

\begin{figure}[t]
\begin{center}
\epsfig{file=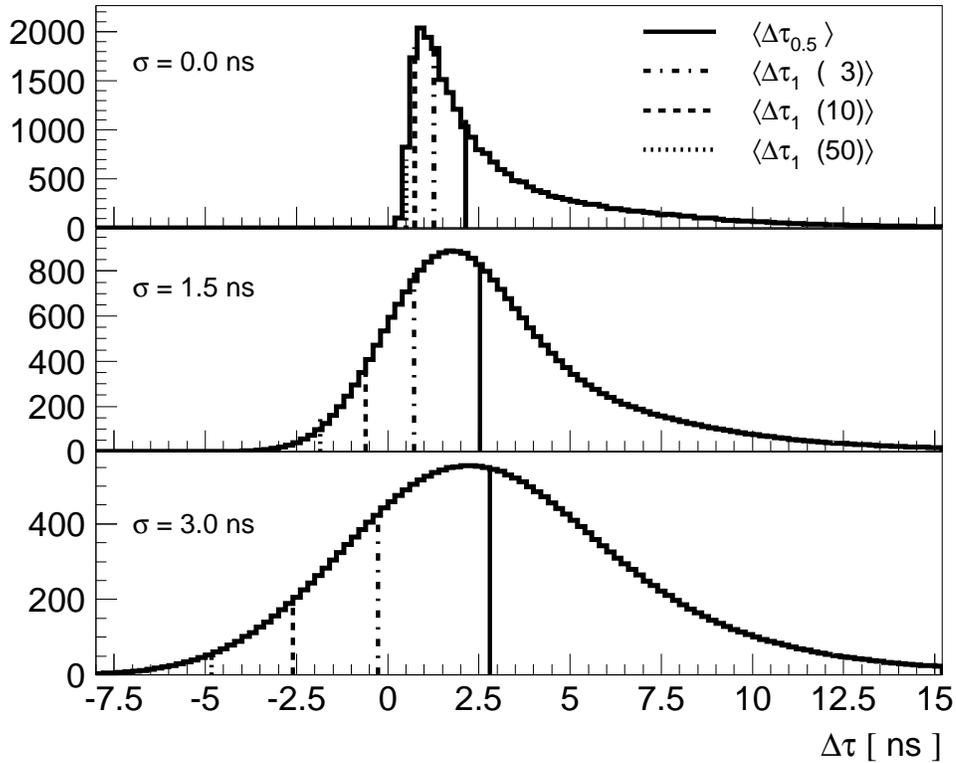, width=15.0cm}
\caption{Muon arrival time distributions $\tau_{1}(n)-\tau_{c}$ 
with the multiplicity dependence of $\langle \Delta\tau_{1}(n)\rangle$
and the influence of the time resolution. The shown distributions 
stem from the accumulation of 100 simulated Fe and 100 simulated 
proton induced EAS (zenith angle of incidence 0$^\circ$) of 
$3\cdot10^{15}$ eV, observed in the range \mbox{$70\, {\rm m} 
\le R_\mu < 80$ m}. In the present case of relatively small observation 
distances from the shower axis differences between Fe and proton induced 
showers would appear completely covered by the discussed fluctuations.}
\end{center}
\end{figure}

With increasing $n$ and in the case of infinite time resolution ($\sigma = 0$ ns) of the timing 
detector
$\langle \Delta\tau_{1}(n)\rangle$ approaches the fictitious arrival time of the shower front 
(which appears for small $R_\mu$ 
and infinite time resolution near $\tau_c$). In addition to the fluctuations of the arrival 
times of
the first muon due to the registered multiplicity (which in practical experimental cases 
involves also the influence of the detector response) there is the influence of the finite time 
resolution. It broadens the observed distributions and smears out the original asymmetry. As 
an example also the expectation for the (global) median $\langle \Delta\tau_{0.5}\rangle$ is 
shown. The averaged global median does not depend much on the time resolution. In contrast 
the local quantities underly the trends of $\langle \Delta\tau_{1}(n)\rangle$, i.e. they increase 
with the multiplicity and with the time resolution. 

\begin{figure}[t]
\begin{center}
\epsfig{file=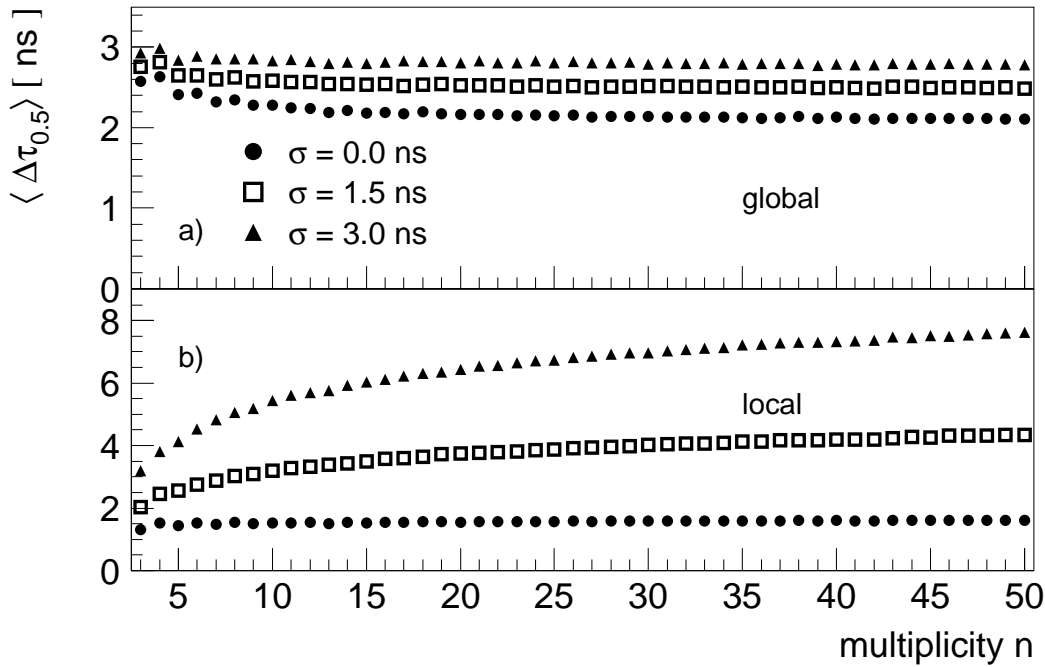, width=16.0cm}
\caption{Multiplicity and time resolution dependence for the mean values 
of global and local time parameters of single muon arrival time distributions.
}
\end{center}
\end{figure}

These features are illustrated in Fig.~3 and lead to local shower 
profiles which are distinctly influenced by the detector qualities 
(time resolution and response for the muon multiplicity i.e. the detector area)
as well as by the natural shower fluctuations.
The influence is very pronounced in the central region 
of the shower with the largest multiplicities. 
As examples Fig.~4 displays local time profiles $\langle \Delta\tau_{0.5}\rangle$ for proton induced
showers observed with different time resolution. The profiles approach the ideal case
at larger distances from the core \mbox{($>$ approx.\,250 m)}, where the time resolution loses
obviously the influence.
It should be noted that {\it local} time profiles do exhibit nearly insignificant
differences for different masses of the EAS primaries. This feature is contrast to observations
of {\it global} time parameters [15].\\

\begin{figure}[bt]
\begin{center}
\epsfig{file=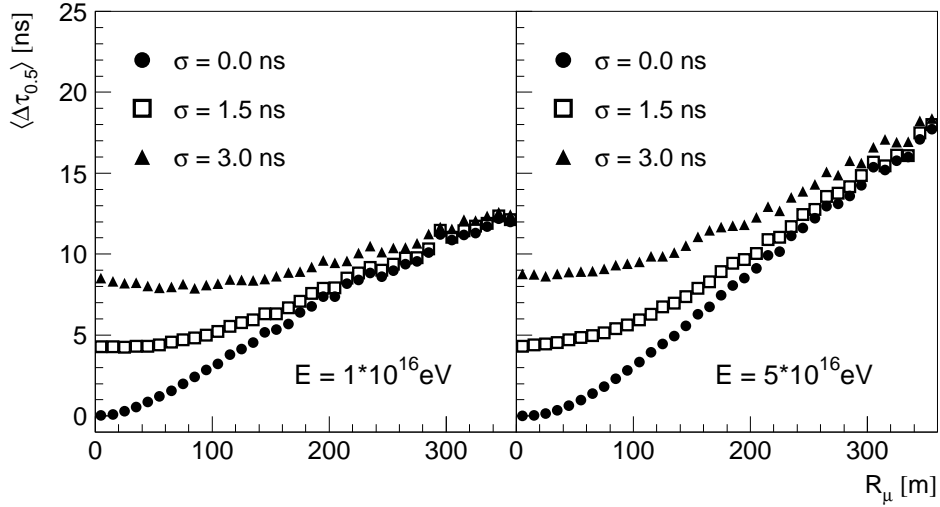, width=14.0cm}
\caption{Local time profiles of proton induced 
EAS for different primary energies and
different time resolution, virtually registered with a 
detector eye of 400 detectors of 0.5~m$^2$
(approximately like the timing facility of the 
KASCADE central detector [12]).
}
\end{center}
\end{figure}


\section{How to account for the dependence on the registered muon multiplicity}

In actual experimental observations of muon arrival time distributions like in the KASCADE
experiment the profiles are derived from EAS events, including all different multiplicities (n 
$>$ n$_s$), 
as registered by the timing detectors. Due to the lateral distribution of the EAS muon
component the average (registered) multiplicities in an EAS are depending on $R_\mu$. They
depend also on the type and energy of the primary, and even if the energy is approximately
specified, the observed time profiles originate from a superposition of various multiplicities,
varying  with the distance from the shower core. That feature leads to interference effects,
distorting the predicted quasi-parabolic shape of the time profile. Fig.~5, which 
displays measured EAS time profiles [26], exhibits this effect.

\begin{figure}[t]
\begin{center}
\epsfig{file=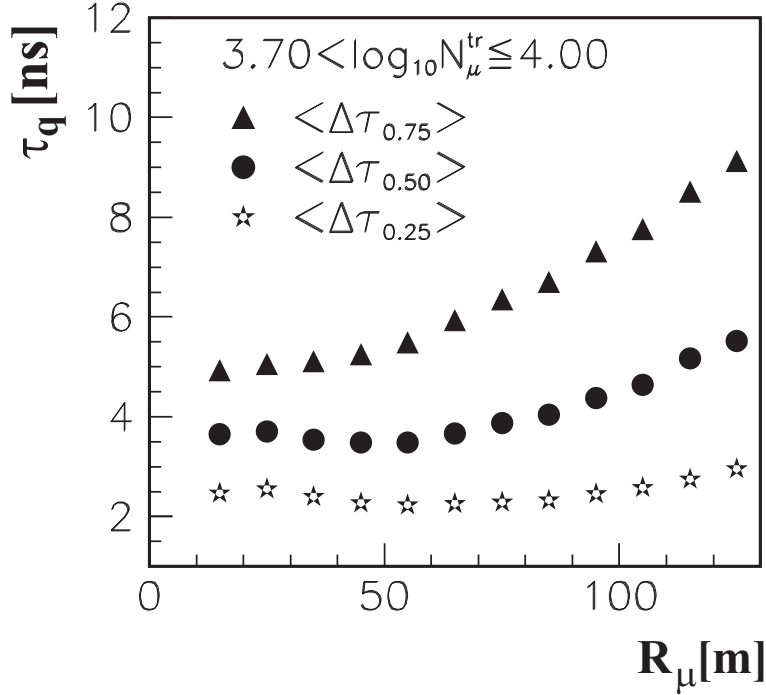, width=10.0cm}
\caption{The observed variation of the mean values of the median, 
first and third quartile distributions with the
distance from the shower axis (from [26]), 
extracted without multiplicity correction for an EAS sample within 
a particular range of $\log_{10}N^{tr}_\mu$ (indicating [27] the 
primary energy range of approx.
$1.6\cdot 10^{15}\,{\rm eV}$ to $3.2\cdot 10^{15}\,{\rm eV}$).
}
\end{center}
\end{figure}

In principle, it would be desirable to extract the shower profile 
for each multiplicity separately. However, this attempt would meet 
serious problems of the statistical accuracy of the results. 
In addition such a procedure would lead to a less transparent
selection of shower events. The main problem, however is that the 
appearance of the distortions depends on the response qualities of 
the particular detector arrangement, so that measurements by
different detector arrays are hardly directly comparable.
There are various ways to approach a representative result about 
the EAS time structure from measurements of local quantities.

\begin{figure}[t]
\begin{center}
\epsfig{file=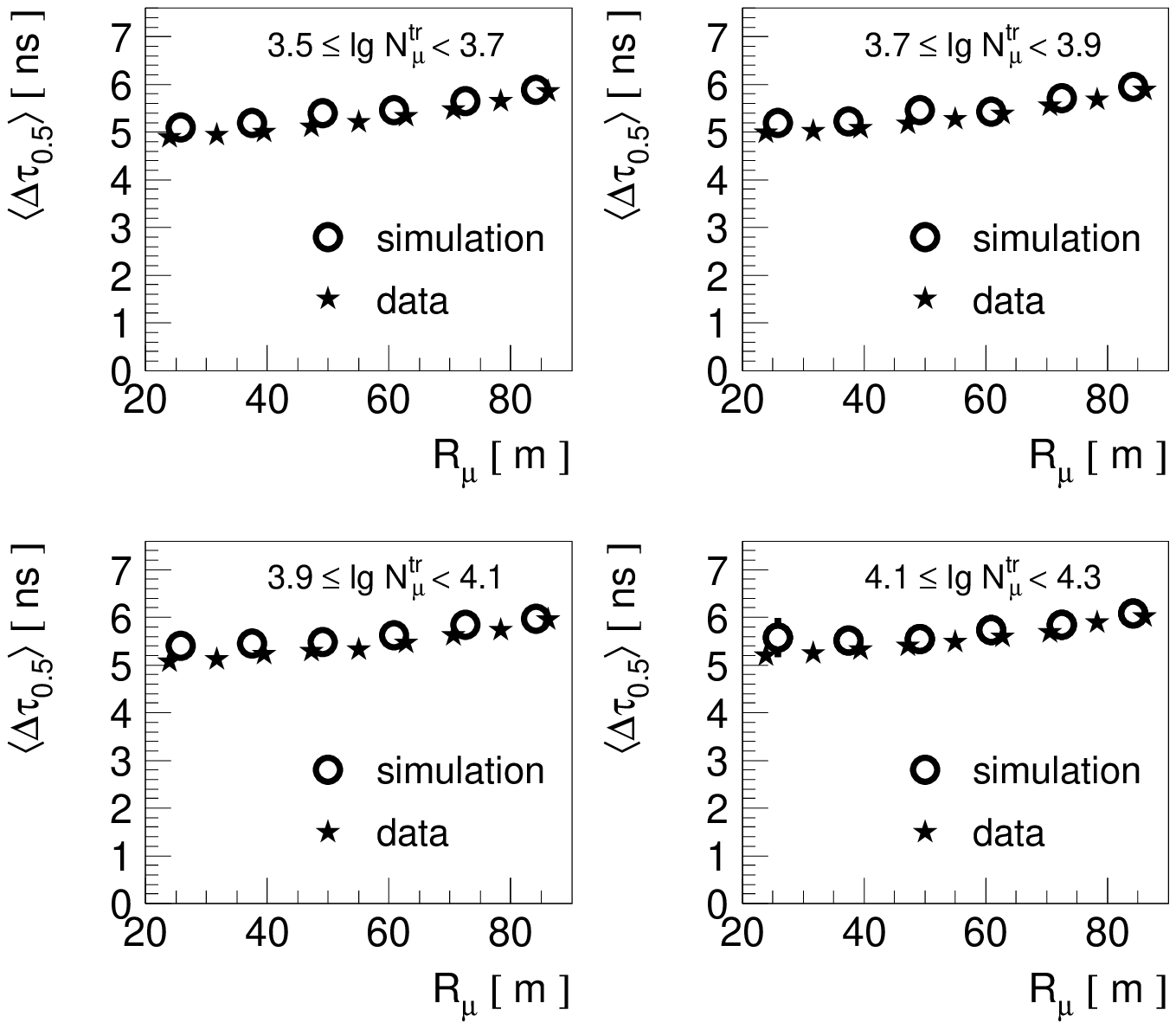, width=15.0cm}
\caption{Local EAS shower profiles ($\langle \Delta\tau_{0.5}\rangle$), 
corrected for the multiplicity dependence, as compared with 
predictions of EAS simulations using the Monte Carlo code CORSIKA 
[23] and taking into account the time resolution, 
for different EAS muon sizes (corresponding to an energy range 
from $6\cdot10^{14}$ eV to $6.3\cdot10^{15}$ eV).}
\end{center}
\end{figure}
 
\begin{itemize}
 
\item The standard concept to compare experimental results with 
theoretical predictions are procedures simulating the experimental 
conditions and response folded with the predictions
and comparing the resulting distributions with the measured results. 
This will give an impression about agreement or disagreement and on 
the undistorted time profiles, but does not immediately enable the 
comparison with other experiments of different (often 
unknown) quality and conditions.
 
\item The measured time parameters, deduced for each event from the 
single muon arrival time distributions registered with varying 
multiplicity, get scaled to a chosen reference value of
the multiplicity by a correction according to predictions by 
simulation calculations (indicated for the case of the mean 
value $\Delta \tau_{mean}$ by a calibration curve in Fig.~3). The
appearance of time profiles depends on the choice of the reference 
multiplicity. The correction procedure needs detailed simulation 
calculations of muon arrival time distributions of the kind shown 
in Fig.~3. The procedure has been successfully 
applied in recent KASCADE experiments [17,19].
 
\item The observation that for global time quantities the influences 
of the multiplicity and the time resolution are less pronounced, 
suggests to relate the muon arrival times to the arrival
time of the shower centre by simulating the time difference between 
the arrival time $\tau_1$ of the local first muon and the 
arrival $\tau_c$ of the EAS core. In this way [20] the local quantities 
are transformed into pseudo-global time parameters, which display the
EAS time structure rather realistically, but invoke EAS simulations, 
specified in detail (see also ref. [11]). While only the shape of the 
arrival time distributions enters in the above multiplicity
calibration procedure, the transformation to pseudo-global quantities 
stresses also the absolute time difference $\tau_1-\tau_c$ between the arrival
of the first muon and of the shower core in the plane perpendicular to the axis.
 
\end{itemize}

Finally Fig.~6 presents a result of an experimental investigation [19]  
of EAS  time profiles using the KASCADE detector. For different ranges 
of the truncated muon number $N^{tr}_\mu$ (used as approximate energy 
estimator [27]) and with a consistent correction for the multiplicity
dependence, the experimental results are compared with CORSIKA [23] 
simulations, adopting a mass composition p:O:Fe = 1:1:1.


\section{Concluding remarks}

The shape of measured muon arrival time distributions, which refer to 
the arrival time $\Delta\tau_{1}$ of the first locally registered muon, 
experiences some distortions which arise from the locally varying 
superposition of different muon multiplicities (spanning the observed 
individual event distribution). The effect of multiplicity fluctuations 
is entangled and amplified in an intricate way with the response and time 
resolution of the detector setup. Due to the larger intensity of the 
electromagnetic EAS component and the smaller fluctuations of the local 
arrival time of the first {\it charged} particle, the effects are less 
pronounced in practical cases of studies of the time structure of the 
charged particle component [12]. In principle, however, the dependence of 
$\langle \Delta\tau_{1}\rangle$ from the observed multiplicity is also 
present in simulation studies  with an ideal time resolution, though 
mostly covered by the assumption of large collection areas of the ideal 
detectors. There are various procedures, which remedy the distortions 
and reveal the basic experimental results of the structure of the EAS disk for a reasonable 
comparison with theoretical predictions or other experiments. The described 
procedures are weakly dependent on the used interaction model, 
governing the EAS development, and on the adopted mass composition. 
They invoke explicitely EAS Monte Carlo simulations. The described distortions 
of the observed muon arrival time distributions lose their significance with 
increasing distance $R_\mu$ from the shower axis ( $>$ 250 m)  and with 
increasing  energy. Nevertheless local time distributions (referring to the 
arrival time of the first local muon) turn out to be rather insensitive to the 
mass of primary cosmic rays. This is in contrast of global arrival times 
[12-15] which refer e.g. to the arrival time of the shower core and measure 
additionally the curvature of the shower disk.\\

\vspace*{0.4cm}
{\noindent \bf Acknowledgements} \\
 
{\it The consideration are borne out from the experimental 
studies of the EAS temporal structure by the KASCADE experiment. 
We acknowledge in particular the clarifying discussions with 
Dr. Andreas Haungs and Dr. Markus Roth. 
Two of us (A.F.B. and I.M.B.) would like to thank for the personal 
support by the WTZ project (RUM 99-005) of the scientific-technical
cooperation understanding between Germany and Romania.\\} 


\end{document}